\documentclass[12pt]{JHEP3}

\usepackage{amsmath}

\newcommand{\be}[1]{ \begin{equation}\label{#1} }
\newcommand{\ee}{\end{equation}}
\newcommand{\ben}[1]{\begin{eqnarray}\label{#1} }
\newcommand{\een}{\end{eqnarray}}

\newcommand{\p}{\partial}
\newcommand{\refb}[1]{(\ref{#1})}

\newcommand{\FF}{{\cal F}}

\newcommand{\DD}{{\cal D}}

\newcommand{\WW}{\cal{W}}

\title{One loop partition function for \\ Topologically Massive Higher Spin Gravity}

\author{
Arjun Bagchi$^1$\footnote{arjun DOT bagchi AT ed DOT ac DOT uk}, Shailesh Lal$^2$\footnote{shailesh AT hri DOT res DOT in}, Arunabha Saha$^2$\footnote{arunabha AT hri DOT res DOT in}, Bindusar Sahoo$^3$\footnote{bsahoo AT ictp DOT it}\\

$^1$ $\,$School of Mathematics, \\
$\;$ $\,$University of Edinburgh \\
$\;$ $\,$Edinburgh EH9 3JZ, UK. \\

$\;$ $^2$Harish-Chandra Research Institute,\\
$\;$ $\,$Chhatnag Road, Jhusi,\\
$\;$ $\,$Allahabad 211019, India.\\

$\;$ $^3$ICTP, High Energy, Cosmology and Astroparticle Physics,\\
$\;$ $\,$Strada Costiera 11, 34151, Trieste, Italy.\\

}

\abstract{We calculate the one loop partition function for topologically massive higher spin gravity (TMHSG) for arbitrary spin by taking the  spin-$3$ TMHSG action constructed in arXiv:1107.0915 and subsequently generalising it for an arbitrary spin. We find that the final result can be put into a product form which cannot be holomorphically factorized giving strong evidence that the topologically massive higher spin gravity is dual to a high spin extension of logarithmic CFT rather than a chiral one.}

\preprint{ EMPG-11-21\\HRI/ST/1110}

\begin{document}

\baselineskip 3.5ex

\section{Introduction}
Possibly the most viable testing ground for theories of quantum gravity has been three dimensional 
gravity in Anti de-Sitter space. Pure Anti de-Sitter gravity in three dimensions, as is well known, has no local propagating degrees of freedom. This feature can be remedied by adding a topological, gravitational Chern-Simons term to the action and then one can show that the linearised equations of motion become that of a massive scalar field. The theory goes under the name of Topologically Massive Gravity (TMG) \cite{{Deser:1981wh, Deser:1982vy}}. 

Through the celebrated AdS/CFT conjecture, there have been attempts to find the dual of the 3d AdS theory in terms of a conformal field theory. The field theory dual to the bulk Einstein theory in $AdS_3$ was initially conjectured to be an extremal CFT \cite{Witten:2007kt}. But a partition function computation by taking into account contributions from classical geometries and also including quantum corrections \cite{Maloney:2007ud} showed that the expected holomorphic factorisation does not hold and there were other studies \cite{Gaberdiel:2007ve} which indicated that this conjecture was incorrect. 

Motivated by \cite{Witten:2007kt}, the authors of \cite{Li:2008dq} looked at Einstein gravity in AdS, now modified by a gravitational Chern-Simons term, with a hope to obtain holomorphic factorization of the partition function. A study of asymptotic symmetries of pure AdS in \cite{BH} had lead to the construction of two copies of the Virasoro algebra with central charges $c_{\pm}= 3\ell/2G$ (where $\ell$ is the AdS radius and $G$ the Newton's constant in three dimensions). With the addition of the Chern-Simons term, the two copies of the Virasoro algebra still emerge as the asymptotic symmetry, but now with modified central charges $c_{\pm} = {3l \over 2G} (1 \mp {1\over \mu\ell}) $, where $1/ \mu$ is the coefficient of the gravitational Chern-Simons term in the action as shown in \cite{Solodukhin:2005ah,Kraus:2005zm}. At the `chiral' point $\mu\ell =1$, one of the central charges vanishes and this led to the Chiral Gravity conjecture in \cite{Li:2008dq}. The authors claimed that at this value of the coupling, the boundary theory dual to TMG in $AdS_3$ was a right-moving chiral CFT with a holomorphically factorizable partition function.

This conjecture was soon hotly debated (see, for example, the works \cite{confusion}), and in \cite{Grumiller:2008qz,Grumiller:2009mw} it was shown that TMG at the chiral point was more generally dual to a Logarithmic Conformal Field Theory (LCFT) and that there were solutions which carried negative energy at the chiral point, thus invalidating the earlier conjecture. A more complete analysis in terms of holographic renormalisation techniques \cite{Skenderis:2009nt} was carried out and the results supported the latter claim of the duality to LCFT. One of the more robust checks of this conjecture was the recent computation of the one-loop partition function \cite{Gaberdiel:2010xv}. It was conclusively shown that there is no holomorphic factorisation of the one-loop partition function at the chiral point. The structure of the gravity calculation also matched with expectations from LCFT. The authors \cite{Gaberdiel:2010xv} found an exact matching with a part of the answer, $viz.$ the single-particle excitations and offered substantial numerical evidence of the matching of the full partition functions. 

Recently, there has been a renewed interest in theories of massless fields with spin greater than two in AdS spacetimes. It is known that these interacting higher spin theories are not sensible in flat spacetimes. Even in AdS, such theories generically require an infinite tower of fields with all possible spins to be consistent. Remarkably, in three dimensions, there can be a truncation to fields with spin less than and equal to $N$ for any $N$. It has been argued in \cite{Campoleoni:2010zq,Henneaux:2010xg} on the basis of a Brown-Henneaux analysis that classically, these theories have an extended classical ${\WW}_N$ asymptotic symmetry algebra (see also the recent work \cite{Campoleoni:2011hg}). A one-loop computation in {\cite{Gaberdiel:2010ar}}, using the techniques developed in \cite{David:2009xg}, showed that this symmetry is indeed perturbatively realised at the quantum level as well. This was an important ingredient in formulating the duality \cite{Gaberdiel:2010pz} between higher-spin theories and ${\WW}_N$ minimal models in the large-$N$ limit. There was also a subsequent work \cite{Castro:2010ce}, in which the authors provided a bound on the amount of higher spin gauge symmetry present. This was regarded as a “gravitational exclusion principle,” where quantum gravitational effects place an upper bound on the number of light states in the theory. Different tests of this duality have been performed successfully subsequent to its proposal, see \cite{other}.

The higher spin theories described above, like Einstein AdS gravity in three dimensions, do not have any propagating degrees of freedom. It is natural thus to ask if one can generalise Topologically Massive Gravity to theories of higher spin. We, in \cite{Bagchi:2011vr}, initiated a construction of this theory, which we call Topologically Massive Higher Spin Gravity (TMHSG) (see also the overlapping work \cite{Chen:2011vp}). As a first step to this end, we studied the quadratic action for a spin-$3$ field in the linearised approximation about $AdS$ and obtained hints that the spin-$3$ theory at the so-called chiral point is dual to a logarithmic CFT. Specifically, we found that the space of solutions developed an extra logarithmic branch at the chiral point. In this paper, we shall perform a quantum test of this conjecture. In particular, we shall compute the one-loop partition function of the spin-$3$ theory on a thermal quotient of $AdS_3$ and show that it does not factorize holomorphically at the chiral point. Our analysis is along the lines of \cite{Gaberdiel:2010xv}, and our results may be viewed as a higher-spin generalisation of theirs. We shall also compute the one-loop partition function for a spin-$N$ generalisation of the action proposed in \cite{Bagchi:2011vr} and show that this property continues to hold. We interpret these results as an indication that the dual CFT at the chiral point is not chiral, and that the results are consistent with the expectation of a high spin extension of a dual logarithmic CFT.

A brief overview of this paper is as follows. In section \ref{bs} we shall review the basic setup for the spin-$3$ calculation. We find that the one-loop partition function receives contributions from transverse traceless spin-$3$, spin-$2$ (coming from the ghost determinant) and transverse spin-$1$ modes (which is the trace of the spin- $3$ field).\footnote{We remind the reader that in \cite{Bagchi:2011vr} we had obtained non-gauge spin-$1$ excitations in the spectrum of the theory, which were not present in the undeformed theory. It is perhaps not surprising that we find an extra spin-$1$ contribution to the one-loop partition function of the theory as compared to that for the ``massless" theory calculated in \cite{Gaberdiel:2010ar}.} We compute the relevant one-loop determinants in section \ref{spin3det}. We find that the spin-$3$ and spin-$1$ contributions referred to above contain terms (apart from the usual holomorphic ones inherited from the undeformed theory) which are non-holomorphic in $q\equiv e^{i\tau}$, where $\tau$ is the modular parameter on the boundary torus of thermal $AdS_3$. In particular, we find the contribution to the one loop partition function at the chiral point, from the spin-$3$ fields to be
\begin{equation}
\label{1.1}
Z_{TMHSG}^{(3)}=\prod_{n=3}^{\infty}{1\over {\left|1-q^{n}\right|^{2}}}\prod_{m=3}^{\infty}\prod_{\bar{m}=0}^{\infty}{1\over\left(1-q^m\bar{q}^{\bar{m}} \right)}\prod_{k=4}^{\infty}\prod_{\bar{k}=3}^{\infty}{1\over\left(1-q^k\bar{q}^{\bar{k}} \right)}.
\end{equation}
The first term is the holomorphic contribution determined from a study of the massless theory in \cite{Gaberdiel:2010ar}, the other terms are non-holomorphic, and new. The middle term is the contribution from the transverse traceless spin-$3$ determinant, while the last term is the transverse spin-$1$ contribution coming from the trace of the spin-$3$ field.

As we can see from the spin-$3$ result that the one loop partition function is equivalent to the spectrum of the linearised equations of motion obtained by us in \cite{Bagchi:2011vr}. The contribution from the traceless spin-$3$ part to the partition function is from $(m,\bar{m})=(3,0)$ onwards which corresponds to the weights of the traceless primaries and its descendants as found in \cite{Bagchi:2011vr} and the contribution from the spin-$1$ trace is from $(k,\bar{k})=(4,3)$ onwards which corresponds to the weights of the trace primaries and its descendants as we found in \cite{Bagchi:2011vr}. Since the one loop calculation can be viewed as the partition function $Tr(q^{L_0}\bar{q}^{\bar{L_0}})$, in general,  we must expect all physical modes that we saw in our classical analysis in \cite{Bagchi:2011vr} to show up in the one loop calculation with exactly the same weights. And this is what we get and hence our classical calculation in \cite{Bagchi:2011vr} and one loop calculations in the present paper are mutually consistent. \footnote{We thank Rajesh Gopakumar for mentioning this point to us.}

In the later part of the paper we discuss the case of general spins in section \ref{arbspin}, where we find analogous results. The relevant excitations are transverse traceless spin-$s$, spin-$s-1$ (coming from the ghost determinant), transverse traceless spin-$s-2$ ones (which is the trace of the spin-$s$ field), transverse traceless spin-$s-3$ (coming from the longitudinal component of trace), transverse traceless spin-$s-4$ (coming from the longitudinal component of the longitudinal component of the trace) and so on upto transverse traceless spin-$1$, of which the spin-$s$, $s-2$, $s-3$, $\cdots$, $1$ contribute non-holomorphically to the one-loop partition function. There are no other relevant excitations coming from the spin-$s$ analysis because of the double-tracelessness condition reviewed, for example, in \cite{Campoleoni:2010zq}. In section \ref{conclusion}, we conclude with a brief interpretation of our results. We will also do a classical analysis for arbitrary spins in appendix \ref{class} and show that the contributions to the partition function also appear in the classical spectrum.

\section{The basic set up for $s=3$} \label{bs}
In this section we will compute the one loop partition function for spin $3$ TMHSG, and in the process build up a mechanism to generalise our calculations to arbitrary spin in the subsequent section. Following the method adopted in \cite{Gaberdiel:2010ar}, we shall compute the one-loop partition function in the Euclideanised version of theory \textit{via} the path integral
\begin{equation}
\label{pathint}
Z^{\left(s\right)}=\frac{1}{\text{Vol(gauge group)}}\int \left[D\phi_{\left(s\right)}\right]e^{-S\left[\phi_{\left(s\right)}\right]}.
\end{equation}
In the one-loop approximation, only the quadratic part of the action $S\left[\phi_{\left(s\right)}\right]$ is relevant. This has been worked out for TMHSG, for the case of $s=3$, in \cite{Bagchi:2011vr}. In the Euclidean signature it takes the form  
\be{1}
S={1\over2}\int d^3 x ~ \sqrt{g} \phi^{MNP} \left[\hat{\FF}_{MNP}-{1\over 2}\hat{\FF}_{(M}g_{NP)}\right],
\ee
where
\be{1.1a}
\hat{\FF}_{MNP} = \DD^{(M)}\FF_{MNP}\equiv \FF_{MNP}+{i\over 6\mu}\varepsilon_{QR(M}\nabla^{Q}\FF^{R}_{~NP)},
\ee
and
\be{1.1b}
\FF_{MNP} = \Delta\phi_{MNP}-\nabla_{(M|}\nabla^{Q}\phi_{|NP)Q}+{1\over 2}\nabla_{(M}\nabla_{N}\phi_{P)}-{2\over \ell^2}g_{(MN}\phi_{P)}.
\ee
As always, the brackets ``( )" denote the sum of the minimum number of terms necessary to achieve complete symmetrisation in the enclosed indices without any normalisation factor. Let us also define the operation of $\DD^{(M)}$ on the trace of the spin-3 field by taking the trace of the expression in \refb{1.1a}
\be{1.3}
\DD^{(M)}\phi_{M}=\phi_{M}+{i\over 6\mu}\varepsilon_{QRM}\nabla^{Q}\phi^{R}
\ee
We now decompose the fluctuations $\phi_{MNP}$ into transverse traceless $\phi^{(TT)}$, trace $\tilde{\phi}_{M}$ and longitudinal parts $\nabla_{(M}\xi_{NP)}$ as\footnote{Note that this decomposition is not orthogonal, in the sense that the trace part contains longitudinal terms, and the longitudinal part is not traceless. See \cite{Gaberdiel:2010ar}.}
\be{1.2}
\phi_{MNP}=\phi^{(TT)}_{MNP}+\tilde{\phi}_{(M}g_{NP)}+\nabla_{(M}\xi_{NP)}.
\ee
Following \cite{Gaberdiel:2010ar}, we use gauge invariance, and orthogonality of the first two terms in \refb{1.2} to decompose the action \refb{1} as
\be{1.4}
S[\phi_{MNP}]=S[\phi^{(TT)}_{MNP}]+S[\tilde{\phi}_{M}],
\ee
where
\be{1.5a}
S[\phi^{(TT)}_{MNP}] = -{1\over 2}\int d^3 x \sqrt{g}\phi^{(TT)MNP}\left[-\DD^{(M)}\Delta\right] \phi^{(TT)}_{MNP},
\ee
and
\be{1.5b}
S[\tilde{\phi}_{M}] = {9\over 4}\int d^3 x ~ \sqrt{g}\left[8\phi^{M} \DD^{(M)}\left(-\Delta+{7\over \ell^2}\right)\phi_{M}-\phi^{M}\DD^{(M)}\nabla_{M}\nabla^{Q}\phi_{Q}\right].
\ee
Now one can further decompose $\tilde{\phi}_{M}$ into its transverse and longitudinal parts as
\be{1.6}
\tilde{\phi}_{M}=\tilde{\phi}^{(T)}_{M}+\nabla_{M}\chi.
\ee
The action for $\phi_{M}$ then becomes
\be{1.7}
S[\tilde{\phi}_{M}] = {9\over 4}\int d^3 x ~ \sqrt{g}\left[8\tilde{\phi}^{(T)M} \DD^{(M)}\left(-\Delta+{7\over \ell^2}\right)\tilde{\phi}^{(T)}_{M}+9\chi\left(-\Delta+{8\over \ell^2}\right)(-\Delta)\chi\right].
\ee
 We see that the $\chi$ part of the action is same as that obtained for the massless theory in \cite{Gaberdiel:2010ar} and will subsequently cancel with the relevant term coming from the ghost determinant, which arises from making the change of variables \refb{1.2} in the path integral \refb{pathint}. The ghost determinant-- being independent of the structure of the action-- is essentially be the same as that obtained \cite{Gaberdiel:2010ar}, see their expression $(2.14)$. This turns out to imply that although that the trace contribution does not cancel with the ghost determinant, the contribution coming from its longitudinal part does. This is in accordance with the observation in \cite{Bagchi:2011vr} that the trace in TMHSG is not pure gauge unlike the massless theory. It is however still true even in the topologically massive theory that $\nabla^{M}\phi_{M}$ is pure gauge 
and hence the longitudinal contribution from the trace $\phi_M$ does cancel with the ghost determinant. 

The spin-$3$ contribution to the one loop partition function for TMHSG is given by 
\begin{eqnarray}\label{1.71}
 Z_{TMGHS}^{(3)}=Z_{gh}^{(3)}&\times&(\det[(-\DD_{(M)}\Delta)]^{TT}_{(3)})^{-\frac{1}{2}} \times (\det[\DD_{(M)}(-\Delta+\frac{7}{\ell^2})]^{T}_{(1)})^{-\frac{1}{2}}\nonumber\\ &\times& (\det[-\Delta(-\Delta+\frac{8}{\ell^2})]_{(0)})^{-\frac{1}{2}},
\end{eqnarray}
where $Z_{gh}^{(3)}$ is the ghost determinant arising as a Jacobian factor corresponding to the change of variables \refb{1.2}. The ghost determinant has been obtained in \cite{Gaberdiel:2010ar}, see their expression $(3.9)$, and is given by
\be{1.72}
 Z_{gh}^{(3)}=[\det(-\Delta+\frac{6}{\ell^2})^{TT}_{(2)} \times \det(-\Delta+\frac{7}{\ell^2})^{T}_{(1)} \times \det (-\Delta+\frac{8}{\ell^2})_{(0)}]^{\frac{1}{2}}.
\ee
Therefore, the spin-$3$ contribution to the one loop partition function is given by
\ben{1.73}
 Z_{TMGHS}^{(3)} &=& [\det[-\DD_{(M)}\Delta]^{TT}_{(3)}]^{-\frac{1}{2}}[\det[\DD_{(M)}]^{T}_{(1)}]^{-\frac{1}{2}}[\det[-\Delta+\frac{6}{\ell^2})]^{TT}_{(2)}]^{\frac{1}{2}} \nonumber \\
&\equiv& Z_{massless}^{(3)} Z_{M}^{(3)},
\een
where $Z_{massless}^{(3)}$ is the massless spin-3 partition function obtained in \cite{Gaberdiel:2010ar}, which is 
\be{1.74}
Z_{massless}^{(3)}=\prod_{s=3}^{\infty}{1\over{\left|1-q^{n}\right|}^{2}}.
\ee
It remains to determine $Z_{M}^{(3)}$. In order to do so, we shall follow the method of \cite{Gaberdiel:2010xv}, and first calculate the absolute value $\vert Z_{M}^{(3)}\vert$. Following \cite{Gaberdiel:2010xv}, we define
\be{1.74}
 |Z_{M}^{(3)}|=\left[\det(\DD_{(M)}\bar{\DD}_{(M)})^{TT}_{(3)}\right]^{-\frac{1}{4}}\times\left[\det(\DD_{(M)}\bar{\DD}_{(M)})^{T}_{(1)}\right]^{-\frac{1}{4}}.
\ee
The subscript $(3)$ and $(1)$ signifies that the operators are acting on transverse traceless spin-$3$ and transverse spin-$1$ respectively. Using \refb{1.1a} and \refb{1.3}, we can show that
\ben{1.75}
 \DD_{(M)}\bar{\DD}_{(M)}\phi^{TT}_{MNP} &=& -\frac{1}{4\mu^2}(\Delta+\frac{4}{\ell^2}-4\mu^2)\phi^{TT}_{MNP} \nonumber \\
 \DD_{(M)}\bar{\DD}_{(M)}\phi_P^{T} &=& -\frac{1}{36\mu^2}[-\Delta+(36\mu^2-\frac{2}{\ell^2})]\phi_{P}^{T}.
\een
Therefore,
\be{1.74}
 |Z_{M}^{(3)}|=\left[\det\left(-\Delta+4(\mu^2-\frac{1}{\ell^2})\right)_{(3)}\right]^{-\frac{1}{4}}\times\left[\det\left(-\Delta+(36\mu^2-\frac{2}{\ell^2})\right)_{(1)}\right]^{-\frac{1}{4}}.
\ee

\section{One-loop determinants for spin-3}
\label{spin3det}
We shall evaluate the one-loop determinants utilising the machinery developed in \cite{David:2009xg}, according to which the relevant determinant takes the following form 
\begin{equation} \label{determ}
 -\log {\det(-\Delta+\frac{m_s^2}{\ell^2})^{TT}_{(s)}}=\int_0^\infty\frac{dt}{t} K^{(s)}(\tau,\bar{\tau};t) e^{-m_{(s)}^2 t},
\end{equation}
where $K^{(s)}$ is the spin-$s$ heat kernel given by
\begin{equation} \label{hk}
 K^{(s)}(\tau,\bar{\tau};t)=\sum_{m=1}^\infty \frac{\tau_2}{\sqrt{4\pi t}|sin\frac{m\tau}{2}|^2}cos(sm\tau_1)e^{-\frac{m^2\tau_2^2}{4t}} e^{-(s+1)t}.
\end{equation}
We therefore obtain
\begin{equation}
\begin{split}
-\log{\det \left(-\Delta+\frac{4(\mu^2l^2-1)}{\ell^2}\right)_{(3)}^{TT}} &=\sum_{m=1}^\infty \frac{1}{m}\frac{cos(3m\tau_1)}{|sin\frac{m\tau}{2}|^2}e^{-2\mu\ell m\tau_2}\\
&= \sum_{m=1}^{\infty}\frac{2}{m} \frac{q^{3m}+\bar{q}^{3m}}{(1-q^m)(1-\bar{q}^m)}(q\bar{q})^{m(\mu\ell -1)}, \\
\end{split}
\end{equation}
and similarly,
\begin{equation}
 -\log{\det \left[-\Delta+\frac{36\mu^2l^2-2}{\ell^2}\right]^{T}_{(1)}}=\sum_{m=1}^{\infty}\frac{2}{m}\frac{q^m+\bar{q}^m}{(1-q^m)(1-\bar{q}^m)}(q\bar{q})^{3 \mu\ell m}.
\end{equation}
We have used the notations and conventions of \cite{David:2009xg} throughout. We remind the reader that $q\equiv e^{i\tau}$, where $\tau=\tau_1+i\tau_2$ is the complex structure modulus on the boundary torus of thermal $AdS_3$.
We can now put together all the contributions into the expression for the one-loop partition function
\ben{partition}
\log |Z_{M}^{(3)}| &=& -{1\over 4}\log{\det \left(-\Delta+\frac{4(\mu^2l^2-1)}{\ell^2}\right)_{(3)}^{TT}} -{1\over 4}\log{\det \left[-\Delta+\frac{36\mu^2l^2-2}{\ell^2}\right]^{T}_{(1)}} \nonumber \\
&=& \sum_{m=1}^{\infty}\frac{1}{2m} \frac{q^{3m}+\bar{q}^{3m}}{(1-q^m)(1-\bar{q}^m)}(q\bar{q})^{m(\mu\ell -1)} + \sum_{m=1}^{\infty}\frac{1}{2m}\frac{q^m+\bar{q}^m}{(1-q^m)(1-\bar{q}^m)}(q\bar{q})^{3 \mu\ell m}. \nonumber \\
\een
Following \cite{Gaberdiel:2010xv}, we infer from this that at the chiral point we have (after rewriting $\log |Z_{M}^{(3)}|= {1\over 2}\log Z_{M}^{(3)} + {1\over 2}\log \bar{Z}_{M}^{(3)}$)
\be{parttitionchiral}
\log Z_{M}^{(3)} =\sum_{m=1}^{\infty} {1\over m} {{\left(q^{3m}+{(q\bar{q})}^{3m}q^m \right)}\over {\left(1-q^m\right)\left(1-\bar{q}^{m}\right)}}.
\ee
Let us now note that 
\ben{p1}
\sum_{n=1}^{\infty} {1\over n} {q^{3n} \over{\left(1-q^n\right)\left(1-\bar{q}^{n}\right)}} &=& -\sum_{m=3}^{\infty}\sum_{\bar{m}=0}^{\infty}\log\left(1-q^m\bar{q}^{\bar{m}} \right) \nonumber \\
\sum_{n=1}^{\infty} {1\over n} {(q\bar{q})^{3n} q^{n}\over{\left(1-q^n\right)\left(1-\bar{q}^{n}\right)}} &=&-\sum_{m=4}^{\infty}\sum_{\bar{m}=3}^{\infty}\log\left(1-q^m\bar{q}^{\bar{m}} \right),
\een
and hence the spin-$3$ contribution to the full partition function at the ciral point, factorizes as (after including the massless contribution from \cite{Gaberdiel:2010ar})
\be{p2}
Z_{TMHSG}^{(3)}=\prod_{n=3}^{\infty}{1\over {\left|1-q^{n}\right|^{2}}}\prod_{m=3}^{\infty}\prod_{\bar{m}=0}^{\infty}{1\over\left(1-q^m\bar{q}^{\bar{m}} \right)}\prod_{k=4}^{\infty}\prod_{\bar{k}=3}^{\infty}{1\over\left(1-q^k\bar{q}^{\bar{k}} \right)}.
\ee
This is the expression \refb{1.1} that we had mentioned in the introduction. As noted there, the partition function does not factorize holomorphically. A chiral CFT would not give rise to such a partition function. We therefore take this as a one-loop clue that the dual CFT to TMHSG is indeed not chiral, but logarithmic. We remind the reader that an analogous result was found in \cite{Gaberdiel:2010xv} for the case of topologically massive gravity, from which the authors were able to conclude that the dual CFT was indeed logarithmic.
\section{Generalisation to arbitrary spin}
\label{arbspin}
In this section, we will consider a natural generalisation of our spin-$3$ action \refb{1} to an arbitrary spin-$s$ field. We will perform the analysis of Sections \ref{bs} and \ref{spin3det} and show that holomorphic factorisation does not occur. We take the action\footnote{In principle, as mentioned in \cite{Bagchi:2011vr}, we can obtain the quadratic action for topologically massive spin-$N$ fields from the Chern-Simons formulation of the theory with unequal levels. However, this action is a natural extension of the spin-$2$ and the spin-$3$ actions, and since the structure is fixed by gauge invariance, the action that follows from the Chern-Simons formulation with unequal levels should give rise to the same action. The Chern-Simons formulation should also fix the normalisation of parity violating C-S term in \refb{arb1} which was fixed by demanding uniqueness of the chiral point.}
\be{arb}
S={1\over2}\int d^3 x ~ \sqrt{g} \phi^{M_1 M_2 \cdots M_s} \left[\hat{\FF}_{M_1 M_2 \cdots M_s}-{1\over 2}\hat{\FF}^{P}_{~P(M_{1}\cdots M_{s-2}}g_{M_{s-1} M_s)}\right],
\ee
where
\be{arb1}
\hat{\FF}_{M_1 \cdots M_s} = \DD^{(M)}{\FF}_{M_1 M_2 \cdots M_s}\equiv {\FF}_{M_1 \cdots M_s}+{i\over s(s-1)\mu}\varepsilon_{QR(M_1}\nabla^{Q}{\FF}^{R}_{~M_2\cdots M_s)},
\ee
and
\ben{arb1.01}
{\FF}_{M_1 M_2 M_3..... M_s} &\equiv & \Delta \phi_{M_1 M_2 ..... M_s}-\nabla_{(M_1|}\nabla^{Q}\phi_{|M_2.....M_s)Q}+{1\over 2}\nabla_{(M_1}\nabla_{M_2}\phi_{M_3 M_4....M_s) P}^{~~~~~~~~~~~~~~P} \nonumber \\
&& -{1\over \ell^2}\left\{\left[s^2-3s\right]\phi_{M_1 M_2....M_s}+2g_{(M_1 M_2}\phi_{M_3 M_4...M_s) P}^{~~~~~~~~~~~~~P}\right\}. \nonumber \\
\een
The normalisation $1\over s(s-1)$ in equation \refb{arb1} is fixed by demanding that even in the spin-$N$ case, the chiral point remains $\mu\ell=1$. One can check that the spin-$s$ generalised action \refb{arb} is invariant under the gauge transformation $\phi_{M_1\cdots M_s}\to \phi_{M_1\cdots M_s} +\nabla_{(M_1}\xi_{M_2\cdots M_s)}$ where $\xi_{M_2\cdots M_s}$ is a symmetric-traceless gauge parameter. 

Now we would like to consider the possibility that the higher-spin theory with spins upto $N$ is also dual to a high spin extension of LCFT. One would then expect that along the lines of what we have seen for spin-$3$, the one-loop partition function again would not factorize holomorphically. We will now do a one-loop analysis using the action \refb{arb} with this goal in mind. An essential ingredient for this analysis -- as in the case of spin 3 -- is the action of $\DD^{(M})$ on symmetric transverse traceless (STT) tensors of rank $s$ and below. 
Using the definition of $\DD^{(M)}$ in \refb{arb1}, one can show that 
\ben{arb2}
\bar{\DD}^{(M)}\DD^{(M)}\phi_{M_1\cdots M_s}^{(TT)} &=& {1\over {(s-1)}^{2}\mu^{2}}\left[-\Delta + {m_{s}^{2}\over \ell^2}\right]\phi_{M_1 \cdots M_s}^{(TT),} \nonumber \\
\bar{\DD}^{(M)}\DD^{(M)}\phi_{M_3\cdots M_s}^{(s-2,TT)} &=& {{(s-2)}^{2}\over {(s-1)}^{2}s^{2}\mu^{2}}\left[-\Delta + {m_{s-2}^{2}\over \ell^2}\right]\phi_{M_1 \cdots M_s}^{(s-2,TT)}, \nonumber \\
\bar{\DD}^{(M)}\DD^{(M)}\chi_{M_{s-m+1}\cdots M_s}^{(s-m,TT)} &=& {{(s-m)}^{2}\over {(s-1)}^{2}s^{2}\mu^{2}}\left[-\Delta + {m_{s-m}^{2}\over \ell^2}\right]\chi^{(s-m,TT)}_{M_{s-m+1} \cdots M_s},
\een
where
\ben{arb3}
m_{s}^{2} &=& \mu^2 \ell^2 {\left(s-1\right)}^{2}-(s+1) \nonumber \\
m_{s-m}^{2} &=& \mu^2 \ell^2 {s^2 {(s-1)}^{2}\over {(s-m)}^{2}}-(s-m+1), \quad s>m\geq 2.
\een
Here $\phi^{(s-2)}$ and $\chi^{(s-n)}$ are STT tensors of rank $s-2$ and $s-n$ respectively, where $n\geq 3$.

Following the same steps as for spin-3 in the previous section, the one loop partition function can be factorized as
\be{arb4}
Z_{TMHSG}^{(s)}=Z_{(massless)}^{(s)} Z_{M}^{(s)},
\ee
The contributions to $Z_{M}^{(s)}$ come from the determinants of $\DD^{(M)}$ acting on $\phi^{(TT)}$, $\phi^{(s-2,TT)}$ and $\chi^{(s-m,TT)}$ (for $s>m\geq3$). This is apparent from \refb{arb2}, which suggests that $\DD^{(M)}$ acts non trivially on the STT modes of spin-1 and higher. The action of $\DD^{(M)}$ stops at $\chi^{(1)}$ as it acts trivially on scalars, being just the identity map.
After a careful analysis, one sees that
\be{arb5}
\log |Z_{M}^{(s)}|=-{1\over 4}\log \det\left(-\Delta + {m_{s}^{2}\over \ell^2}\right)_{(s)}^{(TT)}-{1\over 4}\sum_{m=2}^{s-1}\log \det\left(-\Delta + {m_{s-m}^{2}\over \ell^2}\right)_{(s-m)}^{(TT)}.
\ee
We find, using the spin-$s$ heat kernel \refb{hk} and the expression \refb{determ} for the determinant,
\ben{arb6}
-\log \det\left(-\Delta + {m_{s}^{2}\over \ell^2}\right)_{(s)}^{(TT)} &=& \sum_{m=1}^{\infty}{2\over m} {(q\bar{q})}^{\left[m{(s-1)\over 2}(\mu\ell-1)\right]}{{q^{ms}+\bar{q}^{ms}}\over {\left|1-q^m\right|}^2}, \nonumber \\
-\log \det\left(-\Delta + {m_{s-m}^{2}\over \ell^2}\right)_{(s-m)}^{(TT)} &=& \sum_{n=1}^{\infty}{2\over n} {(q\bar{q})}^{\left[k(s,m,\mu 
\ell)n\right]}~ {{q^{n(s-m)}+\bar{q}^{n(s-m)}}\over {\left|1-q^n\right|}^2},
\een
Where,
\be{arb61}
 k(s,m,\mu\ell) = {{s(s-1)\mu\ell -(s-m)(s-m-1)}\over {2(s-m)}}.
\ee
At the chiral point $\mu\ell=1$, it becomes
\be{arb7}
k(s,m,1)\equiv k(s,m)={{s(s-1)-(s-m)(s-m-1)}\over {2(s-m)}}={{m(2s-m-1)}\over {2(s-m)}},
\ee
and hence at the chiral point the partition function $Z_{M}^{(s)}$ becomes
\be{arb8}
\log Z_{M}^{(s)} =\sum_{n=1}^{\infty} {1\over n} {{\left(q^{sn}+{\sum_{m=2}^{s-1}}{(q\bar{q})}^{k(s,m)n}q^{(s-m)n} \right)}\over {\left(1-q^n\right)\left(1-\bar{q}^{n}\right)}}.
\ee
After using the identities
\ben{arb9}
\sum_{n=1}^{\infty} {1\over n} {q^{sn} \over{\left(1-q^n\right)\left(1-\bar{q}^{n}\right)}} &=& -\sum_{m=s}^{\infty}\sum_{\bar{m}=0}^{\infty}\log\left(1-q^m\bar{q}^{\bar{m}} \right), \nonumber \\
\sum_{n=1}^{\infty} {1\over n} {(q\bar{q})^{k(s,m)n} q^{(s-m)n}\over{\left(1-q^n\right)\left(1-\bar{q}^{n}\right)}} &=&-\sum_{p=r(s,m)}^{\infty}\sum_{\bar{p}=k(s,m)}^{\infty}\log\left(1-q^p \bar{q}^{\bar{p}} \right)
\een
where,
\be{arb91}
 r(s,m) = k(s,m)+s-m={{s(s-1)+(s-m)(s-m+1)}\over {2(s-m)}},
\ee
the contribution to the full partition function from spin-$s$ field at the chiral point, becomes
\be{arb10}
Z_{TMHSG}^{(s)}=\prod_{n=s}^{\infty}{1\over {\left|1-q^{n}\right|^{2}}}\prod_{m=s}^{\infty}\prod_{\bar{m}=0}^{\infty}{1\over\left(1-q^m\bar{q}^{\bar{m}} \right)}\prod_{t=2}^{s-1}\prod_{p=r(s,t)}^{\infty}\prod_{\bar{p}=k(s,t)}^{\infty}{1\over\left(1-q^p\bar{q}^{\bar{p}} \right)}.
\ee
Hence, the full partition function at the chiral point, in a theory with fields of spin $s=3, \cdots N$ in addition to the spin 2 graviton, becomes
\ben{arb10}
Z_{TMHSG} &=&\prod_{s=2}^{N}\left[\prod_{n=s}^{\infty}{1\over {\left|1-q^{n}\right|^{2}}}\prod_{m=s}^{\infty}\prod_{\bar{m}=0}^{\infty}{1\over\left(1-q^m\bar{q}^{\bar{m}} \right)}\right]\times\left[\prod_{s=3}^{N}\prod_{t=2}^{s-1}\prod_{p=r(s,t)}^{\infty}\prod_{\bar{p}=k(s,t)}^{\infty}{1\over\left(1-q^p\bar{q}^{\bar{p}} \right)}\right], \nonumber \\
\text{Where} ~~ &&  k(s,m)={{s(s-1)-(s-m+1)(s-m-1)}\over {2(s-m)}}, \quad r(s,m)= k(s,m)+s-m \nonumber \\
\een
We will see in appendix \ref{class} that $r(s,m)$ and $k(s,m)$ appear respectively as left and right weights of classical left moving primary solution (or massive primary at the chiral point). In particular, we will also see that, $r(s,m,\mu\ell)\equiv k(s,m,\mu\ell)+s-m$, and $k(s,m,\mu\ell)$ will be the weights of massive primary at a generic point. Thus we also see that for generic spins, our one loop computations and classical computations are also mutually consistent.

Note that the first square bracket contribution starts from $s=2$ whereas the second square bracket contribution starts from $s=3$. This is a novel feature of topologically massive higher spin theory and technically this comes from the fact the trace is not a pure gauge. As the spin increases, the factors contributing to the partition function also increases. This, as discussed before, comes from the fact that longitudinal component of trace, longitudinal components of the longitudinal components of the trace and so on are not pure gauge and starts contributing till one hits a scalar, which is a pure gauge and the contribution terminates. 

One might be worried about the fact that $k(s,m)$ and $r(s,m)$ in \refb{arb10} are not integers for generic spin, and this might not be consistent with the periodicity in $\tau$. However one might note that $r(s,m)-k(s,m)=s-m$, which is an integer and hence the periodicity in $\tau$ is not affected because of the following identity
\be{per}
q^{r(s,m)}\bar{q}^{k(s,m)}={\left({q\bar{q}}\right)}^{k(s,m)}q^{s-m}.
\ee

\section{Conclusions} \label{conclusion}
In this paper, motivated by the results of \cite{Bagchi:2011vr}, we computed the one loop partition function for topologically massive higher spin gravity (TMHSG) for spin-$3$ and later generalised it to arbitrary spin. We find that the one loop partition function does not factorize holomorphically giving strong evidence that the dual theory is a high spin extension of LCFT. This was also anticipated in the classical calculation for spin-$3$ TMHSG in \cite{Bagchi:2011vr} as extra logarithmic modes emerged at the chiral point. Although this result might be tantalising, a CFT interpretation of the result would make the proposal of a high spin extension of dual LCFT more concrete. 

We had speculated on the possible realisation of the symmetry algebra in \cite{Bagchi:2011vr}. One of our speculations was that since the classical ${\WW}_3$ algebra at the chiral point seems to become singular due to the presence of the inverse of the central charge in the coefficient of the non-linear term in the $[W,W]$ commutator, we should look at a contraction of this algebra which would effectively reduce this to a Virasoro algebra. While this was a perfectly correct limit, we also suggested that this could be a rather restrictive realisation. Our analysis in this paper shows that in addition to the non-holomorphic factorisation and a contribution from the trace modes, what we obtain in the 1-loop partition function at the chiral point is the vacuum character of the ${\WW}_3$, for the spin-3 case and a similar ${\WW}_N$ vacuum character for the spin-$N$ generalisation. This indicates conclusively that the ${\WW}$-algebra exists in the chiral limit and does not simply reduce to the Virasoro algebra. This also adds strength to another of our speculations in \cite{Bagchi:2011vr}, $viz.$ at the chiral point one needs to consider the quantum version of the ${\WW}_3$, where the problematic coefficient of the non-linear term gets a shift and hence is not singular at the chiral point. 

This also gives rise to the expectation that we would see novel logarithmic behaviour in the $\WW$-algebra and not the Virasoro alone. Another piece of evidence of the emergent logarithmic nature of the $\WW$-algebra at the chiral point comes from looking at the OPE of the $\WW$-algebra. 
Let us concentrate of the ${\WW}_3$ for the moment. The quantum ${\WW}_3$ introduced by Zamolodchikov in \cite{Zamolodchikov:1985wn} contains two generators, the energy-momentum tensor $T(z)$ and $W(z)$, where $W(z)$ is a primary of spin 3 with respect to $T(z)$. The OPE of $W(z)$ with itself is given by
\ben{wope}
W(z) W(w) \sim&& {c/3 \over{(z-w)^6}} + {2T(w) \over {(z-w)^4}} + {\p T(w) \over {(z-w)^3}} \\
 &&+ { 2 \beta \Lambda(w) + {3\over 10} \p^2 T(w) \over {(z-w)^2}} + {\beta \p \Lambda(w) + {1\over 15} \p^3 T(w) \over {(z-w)}}\nonumber
\een
where $ \Lambda(w) = TT(w) - {3\over 10} \p^2 T(w) $ and $\beta = {16\over{5c +22}}$.
From here we can read off the two-point function
\be{wcor}
\langle W(z) W(w) \rangle =  {c/3 \over{(z-w)^6}}
\ee
The left-moving central charge of the ${\WW}_3$ at the chiral point vanishes and this also means that the two-point function of $W(z)$ vanishes here. This is a clear indication of a logarithmic CFT and we expect the following structure:
\ben{wlogcor}
\langle W(z) W(0) \rangle = 0, \quad \langle W(z) w(0,0) \rangle = {b_w \over z^6} \\
\langle w(z, {\bar z}) w(0,0) \rangle = {1\over z^6} \bigg( B_w - b_w \log m^2 {|z|}^2\bigg)
\een
where $w(z, {\bar z})$ is the logarithmic partner of $W(z)$. $b_w$ is a characterising feature of the LCFT, whereas $B_w$ can be got rid of by a field redefinition. It would be very interesting to see if this expected structure actually emerges from a more thorough analysis in the gravitational set-up generalising the relevant computation done for spin-$2$ in \cite{Skenderis:2009nt,Grumiller:2009mw}. It is plausible that given the complicated structure of the 1-loop partition function that we have obtained here with the additional pieces from the novel trace terms, we would get a richer structure in the logarithmic CFT duals to the TMHSG. We leave addressing this issue for future work. 

Before we conclude, we would like to remind the reader of a few important observations that follow from the results in the present paper. The fact that the 1-loop partition function does not holomorphically factorize gives support to the claim that the the dual to TMHSG is a high spin extension of a logarithmic CFT, and not a chiral one. Additionally there is the presence of a non trivial spin one contribution (in case of spin 3 TMHSG) in the one loop partition function. This points to the fact that the trace modes cannot be gauged away, both of which results are in accordance with our analysis of classical solutions in \cite{Bagchi:2011vr}.

\subsection*{Acknowledegments}
We would like to thank Rajesh Gopakumar and Daniel Grumiller for helpful discussions and for their valuable comments on the manuscript. SL and AS would like to thank CHEP, IISc Bangalore and ICTS Bangalore for hospitality during the course of this work.


\appendix
\section{Classical analysis for generic spins}\label{class}
In this appendix, we will study the classical analysis for generic spins and show that $k(s,m, \mu\ell)$ \refb{arb61} and $r(s,m,\mu\ell) \equiv k(s,m, \mu\ell) +s-m$ will appear as weights (right and left respectively) of massive primaries in the linearised spectrum of generic spins.

The classical equation of motion for a generic spin is
\be{a1}
\hat{\FF}_{M_1\cdots M_s} \equiv \DD^{(M)}{\FF}_{M_1 M_2 \cdots M_s}\equiv {\FF}_{M_1 \cdots M_s}+{1\over s(s-1)\mu}\varepsilon_{QR(M_1}\nabla^{Q}{\FF}^{R}_{~M_2\cdots M_s)}=0,
\ee
where
\ben{a2}
{\FF}_{M_1 M_2 M_3..... M_s} &\equiv & \Delta \phi_{M_1 M_2 ..... M_s}-\nabla_{(M_1|}\nabla^{Q}\phi_{|M_2.....M_s)Q}+{1\over 2}\nabla_{(M_1}\nabla_{M_2}\phi_{M_3 M_4....M_s) P}^{~~~~~~~~~~~~~~P} \nonumber \\
&& -{1\over \ell^2}\left\{\left[s^2-3s\right]\phi_{M_1 M_2....M_s}+2g_{(M_1 M_2}\phi_{M_3 M_4...M_s) P}^{~~~~~~~~~~~~~P}\right\}. \nonumber \\
\een
Using the field redefinition and the gauge condition defined below,
\ben{a3}
\phi_{M_1 \cdots M_s} &=& \tilde{\phi}_{M_1 \cdots M_s}-{1\over s^2}g_{(M_1 M_2}\tilde{\phi}^{(s-2)}_{M_3\cdots M_s)}, \nonumber \\
\nabla^{M}\tilde{\phi}_{M M_2 \cdots M_s} &=& {1\over 2}\nabla_{(M_2}\tilde{\phi}^{(s-2)}_{M_3 \cdots M_s)},
\een
we can write
\be{a4}
{\FF}_{M_1 M_2 M_3..... M_s}=\DD^{(L)}\DD^{(R)}\tilde{\phi}_{M_1 \cdots M_s},
\ee
where
\ben{a5}
\DD^{(L)}\tilde{\phi}_{M_1 \cdots M_s} & \equiv & \tilde{\phi}_{M_1 \cdots M_s}+{\ell\over s(s-1)}\varepsilon_{QR(M_1}\nabla^{Q}\tilde{\phi}^{R}_{~M_2\cdots M_s)}, \nonumber\\
\DD^{(R)}\tilde{\phi}_{M_1 \cdots M_s} & \equiv & \tilde{\phi}_{M_1 \cdots M_s}-{\ell\over s(s-1)}\varepsilon_{QR(M_1}\nabla^{Q}\tilde{\phi}^{R}_{~M_2\cdots M_s)}.
\een
One can now check that the equations of motion implies that
\ben{a6}
&& g^{M_1 M_2}\nabla^{M_3}\cdots\nabla^{M_s}\hat{\FF}_{M_1\cdots M_s}=0 \nonumber \\
\implies && g^{M_1 M_2}\nabla^{M_3}\cdots\nabla^{M_s}{\FF}_{M_1\cdots M_s}=0 \nonumber \\
\implies && \nabla^{M_3}\cdots\nabla^{M_s}\tilde{\phi}^{(s-2)}_{M_3\cdots M_s}=0.
\een
Now let us write down the massive branch equation $\DD^{(M)}\tilde{\phi}=0$, by operating on it from the left by $\bar{\DD}^{(M)}$ which is the same as $\DD^{(M)}$ with $\mu$ replaced by $-\mu$. By replacing $\mu\ell= \pm1$ on the solutions obtained here, we can recover the left and right branch solutions. The equations are
\ben{a7}
\Delta \tilde{\phi}_{M_1\cdots M_s}-{m_{s}^{2}\over \ell^2}\tilde{\phi}_{M_1\cdots M_s} &=& {(2s-3)\over 2s^2}\nabla_{(M_1}\nabla_{M_2}\tilde{\phi}^{(s-2)}_{M_3\cdots M_s)} +{1\over \ell^2 s^2}(s^2-s+2)g_{(M_1 M_2}\tilde{\phi}^{(s-2)}_{M_3\cdots M_s)} \nonumber \\
&& +{1\over s^2}g_{(M_1 M_2}\Delta\tilde{\phi}^{(s-2)}_{M_3\cdots M_s)}-{1\over s^2}g_{(M_1 M_2}\nabla_{M_3}\chi_{M_4\cdots M_s)}^{(s-3)}, \nonumber \\
\Delta \tilde{\phi}^{(s-2)}_{M_3\cdots M_s}-{m_{s-2}^{2}\over \ell^2}\tilde{\phi}^{(s-2)}_{M_3\cdots M_s} &=& {(2s-5)\over {(s-2)}^2}\nabla_{(M_3}{\chi}^{(s-3)}_{M_4\cdots M_s)}-{2\over {(s-2)}^2}g_{(M_3 M_4}\chi_{M_5\cdots M_s)}^{(s-4)}, \quad s>2 \nonumber \\
\Delta {\chi}^{(s-3)}_{M_4\cdots M_s}-{m_{s-3}^{2}\over \ell^2}{\chi}^{(s-3)}_{M_4\cdots M_s} &=& {(2s-7)\over {(s-3)}^2}\nabla_{(M_4}{\chi}^{(s-4)}_{M_5\cdots M_s)}-{2\over {(s-3)}^2}g_{(M_4 M_5}\chi_{M_6\cdots M_s)}^{(s-5)} , \quad s>3\nonumber \\
\vdots \nonumber \\
\Delta {\chi}^{(2)}_{MN}-{m_{2}^{2}\over \ell^2}{\chi}^{(2)}_{MN} &=& {3\over 4}\nabla_{(M}{\chi}^{(1)}_{N)}, \nonumber \\
\Delta {\chi}^{(1)}_{M}-{m_{1}^{2}\over \ell^2}{\chi}^{(1)}_{M} &=& 0,
\een
where 
\begin{equation}
\chi^{(s-m)}=\nabla^{M_3}\cdots\nabla^{M_m}\tilde{\phi}^{(s-2)}_{M_3\cdots M_s},
\end{equation} and $m_{s}$ and $m_{s-m}$ are given in \refb{arb3}. The procedure to solve the above equations is same as the spin-$3$ analysis done in \cite{Bagchi:2011vr}. We need to start with the last equation of \refb{a7}, the solution of which would be given by $h-\bar{h}=\pm1$, where $h$ and $\bar{h}$ are the left and right weights. We will put the solution of $\chi^{(1)}$ into the equation of $\chi^{(2)}$ and decompose $\chi^{(2)}$ into two parts, one carrying the weights of $\chi^{(1)}$ and the other would be transverse, the solution of which would be given by $h-\bar{h}=\pm 2$. Following the steps similarly we will decompose $\chi^{(3)}$ into three parts, one carrying the weight of $\chi^{(1)}$, another carrying the weights of transverse $\chi^{(2)}$ and a part which is transverse on its own carrying weight $h-\bar{h}=\pm 3$. In this way we will finally obtain the full solution. We are not interested in obtaining the explicit form of the final solution (although there is no technical obstacle in doing so). We will, however, be interested in obtaining the weights of the different modes present in the final solution. For that it is sufficient (as per the argument above) to analyse the transverse traceless parts of the equations \refb{a7}. For that let us write the Laplacian acting on traceless rank-s tensor as
\be{a8}
\Delta \phi_{M_1\cdots M_s}=\left[-{2\over \ell^2}\left(L^2 +\bar{L}^{2}\right)-{s(s+1)\over \ell^2}\right]\phi_{M_1\cdots M_s}
\ee
Where, $L^2$ and $\bar{L}^{2}$ are the left and right $SL(2,R)$ casimirs of the isometry group of $AdS_3$ (see eq 4.11 and 4.12 of \cite{Bagchi:2011vr}). The eigenvalues of $L^2$ and $\bar{L}^{2}$ are given by $h(1-h)$ and $\bar{h}(1-\bar{h})$, where $h$ and $\bar{h}$ are the left and right weights of the solution. We will also use the fact that for a rank-$s$, transverse traceless primary, $h-\bar{h}=\pm s$. Thus the different modes of equation \refb{a7}, will carry weights $h-\bar{h}=\pm s$, $h-\bar{h}=\pm(s-2)$, $h-\bar{h}=\pm(s-3)$, $\cdots$, $h-\bar{h}=\pm 1$. By using \refb{a7} and \refb{a8}, we obtain the weights of the different modes as
\ben{a9}
 h-\bar{h}=+s : \quad  && h={{s^2\left(\mu\ell+1\right)-s\left(\mu\ell -1\right)}\over {2s}}, \quad \bar{h}={s(s-1)(\mu\ell -1)\over 2s}, \nonumber \\
h-\bar{h}=-s: \quad  && h={s(s-1)(\mu\ell -1)\over 2s}, \quad \bar{h}= {{s^2\left(\mu\ell+1\right)-s\left(\mu\ell -1\right)}\over {2s}}, \nonumber \\
h-\bar{h}=+(s-m): \quad && h={{(s-m+1)(s-m)+s(s-1)\mu\ell}\over 2(s-m)}=r(s,m,\mu\ell),\nonumber \\
&& \bar{h}= {{s(s-1)\mu\ell-(s-m-1)(s-m)}\over 2(s-m)}=k(s,m,\mu\ell), \quad (s-1)\geq m\geq 2,\nonumber \\
h-\bar{h}=-(s-m): \quad && h=k(s,m,\mu\ell), \quad \bar{h}=r(s,m,\mu\ell), \quad (s-1)\geq m\geq 2
\een
The modes with negative $h-\bar{h}$ will not belong to the massive branch (they appear because we had the operation of $\bar{\DD}^{(M)}$ on the original equation of motion). But these modes will become the right branch solution by taking $\mu\ell=1$. The modes with positive $h-\bar{h}$ will belong to the massive branch solution. For $\mu\ell=1$, they will become the left branch solutions. These are the weights that also appear in the one loop partition function at the chiral point \refb{arb10}.



\begin{thebibliography}{999}

\bibitem{Deser:1981wh}
  S.~Deser, R.~Jackiw and S.~Templeton,
  ``Topologically Massive Gauge Theories,''
  Annals Phys.\  {\bf 140}, 372 (1982)
  [Erratum-ibid.\  {\bf 185}, 406 (1988)]
  [Annals Phys.\  {\bf 185}, 406 (1988)]
  [Annals Phys.\  {\bf 281}, 409 (2000)].

\bibitem{Deser:1982vy}
  S.~Deser, R.~Jackiw and S.~Templeton,
  ``Three-Dimensional Massive Gauge Theories,''
  Phys.\ Rev.\ Lett.\  {\bf 48}, 975 (1982).
  

\bibitem{Witten:2007kt}
  E.~Witten,
  ``Three-Dimensional Gravity Revisited,''
  arXiv:0706.3359 [hep-th].
  

\bibitem{Maloney:2007ud}
  A.~Maloney and E.~Witten,
  ``Quantum Gravity Partition Functions in Three Dimensions,''
  JHEP {\bf 1002}, 029 (2010)
  [arXiv:0712.0155 [hep-th]].

\bibitem{Gaberdiel:2007ve}
  M.~R.~Gaberdiel,
  ``Constraints on extremal self-dual CFTs,''
  JHEP {\bf 0711}, 087 (2007)
  [arXiv:0707.4073 [hep-th]].

  M.~R.~Gaberdiel and C.~A.~Keller,
  ``Modular differential equations and null vectors,''
  JHEP {\bf 0809}, 079 (2008)
  [arXiv:0804.0489 [hep-th]].
  

\bibitem{Li:2008dq}
  W.~Li, W.~Song and A.~Strominger,
  ``Chiral Gravity in Three Dimensions,''
  JHEP {\bf 0804}, 082 (2008)
  [arXiv:0801.4566 [hep-th]].
  
\bibitem{BH}
  J.~D.~Brown, M.~Henneaux,
  ``Central Charges in the Canonical Realization of Asymptotic Symmetries: An Example from Three-Dimensional Gravity,''
  Commun.\ Math.\ Phys.\  {\bf 104}, 207-226 (1986).
  

\bibitem{Solodukhin:2005ah}
  S.~N.~Solodukhin,
  ``Holography with gravitational Chern-Simons,''
  Phys.\ Rev.\  D {\bf 74}, 024015 (2006)
  [arXiv:hep-th/0509148].

\bibitem{Kraus:2005zm}
  P.~Kraus and F.~Larsen,
  ``Holographic gravitational anomalies,''
  JHEP {\bf 0601}, 022 (2006)
  [arXiv:hep-th/0508218].
  
\bibitem{confusion}
S.~Carlip, S.~Deser, A.~Waldron, D.~K.~Wise,
  ``Cosmological Topologically Massive Gravitons and Photons,''
  Class.\ Quant.\ Grav.\  {\bf 26}, 075008 (2009).
  [arXiv:0803.3998 [hep-th]];
 
W.~Li, W.~Song, A.~Strominger,
  ``Comment on 'Cosmological Topological Massive Gravitons and Photons',''  
  [arXiv:0805.3101 [hep-th]].
 
  D.~Grumiller, R.~Jackiw, N.~Johansson,
  ``Canonical analysis of cosmological topologically massive gravity at the chiral point,''
  [arXiv:0806.4185 [hep-th]].  
 
  S.~Carlip, S.~Deser, A.~Waldron, D.~K.~Wise,
  ``Topologically Massive AdS Gravity,''
  Phys.\ Lett.\  {\bf B666}, 272-276 (2008).
  [arXiv:0807.0486 [hep-th]].
  
  G.~Giribet, M.~Kleban, M.~Porrati,
  ``Topologically Massive Gravity at the Chiral Point is Not Chiral,''
  JHEP {\bf 0810}, 045 (2008).
  [arXiv:0807.4703 [hep-th]].
  
  A.~Strominger,
  ``A Simple Proof of the Chiral Gravity Conjecture,''
  [arXiv:0808.0506 [hep-th]].
  
\bibitem{Grumiller:2008qz}
 D.~Grumiller, N.~Johansson,
 ``Instability in cosmological topologically massive gravity at the chiral point,''
  JHEP {\bf 0807}, 134 (2008).
  [arXiv:0805.2610 [hep-th]].

  D.~Grumiller and N.~Johansson,
  ``Consistent boundary conditions for cosmological topologically massive
  gravity at the chiral point,''
  Int.\ J.\ Mod.\ Phys.\  D {\bf 17}, 2367 (2009)
  [arXiv:0808.2575 [hep-th]].
  
\bibitem{Grumiller:2009mw}
  D.~Grumiller and I.~Sachs,
  ``AdS (3) / LCFT (2) ---> Correlators in Cosmological Topologically Massive
  Gravity,''
  JHEP {\bf 1003}, 012 (2010)
  [arXiv:0910.5241 [hep-th]].
  
\bibitem{Skenderis:2009nt}
  K.~Skenderis, M.~Taylor, B.~C.~van Rees,
  ``Topologically Massive Gravity and the AdS/CFT Correspondence,''
  JHEP {\bf 0909}, 045 (2009).
  [arXiv:0906.4926 [hep-th]].

K.~Skenderis, M.~Taylor, B.~C.~van Rees,
  ``AdS boundary conditions and the Topologically Massive Gravity/CFT correspondence,''
  [arXiv:0909.5617 [hep-th]].
  
\bibitem{Gaberdiel:2010xv}
  M.~R.~Gaberdiel, D.~Grumiller and D.~Vassilevich,
  ``Graviton 1-loop partition function for 3-dimensional massive gravity,''
  JHEP {\bf 1011}, 094 (2010)
  [arXiv:1007.5189 [hep-th]].
  
\bibitem{Campoleoni:2010zq}
  A.~Campoleoni, S.~Fredenhagen, S.~Pfenninger, S.~Theisen,
  ``Asymptotic symmetries of three-dimensional gravity coupled to higher-spin fields,''
  JHEP {\bf 1011}, 007 (2010).
  [arXiv:1008.4744 [hep-th]].

\bibitem{Campoleoni:2011hg}
  A.~Campoleoni, S.~Fredenhagen and S.~Pfenninger,
  ``Asymptotic W-symmetries in three-dimensional higher-spin gauge theories,''
  arXiv:1107.0290 [hep-th].
  
\bibitem{Henneaux:2010xg}
  M.~Henneaux, S.~-J.~Rey,
  ``Nonlinear $W_{infinity}$ as Asymptotic Symmetry of Three-Dimensional Higher Spin Anti-de Sitter Gravity,''
  JHEP {\bf 1012}, 007 (2010).
  [arXiv:1008.4579 [hep-th]].
  
\bibitem{Gaberdiel:2010ar}
  M.~R.~Gaberdiel, R.~Gopakumar and A.~Saha,
  ``Quantum $W$-symmetry in $AdS_3$,''
  JHEP {\bf 1102}, 004 (2011)
  [arXiv:1009.6087 [hep-th]].
  
\bibitem{David:2009xg}
  J.~R.~David, M.~R.~Gaberdiel and R.~Gopakumar,
  ``The Heat Kernel on AdS(3) and its Applications,''
  JHEP {\bf 1004}, 125 (2010)
  [arXiv:0911.5085 [hep-th]].
    
  
\bibitem{Gaberdiel:2010pz}
  M.~R.~Gaberdiel, R.~Gopakumar,
  ``An $AdS_3$ Dual for Minimal Model CFTs,''
  Phys.\ Rev.\  {\bf D83}, 066007 (2011).
  [arXiv:1011.2986 [hep-th]].

\bibitem{Castro:2010ce}
  A.~Castro, A.~Lepage-Jutier and A.~Maloney,
  ``Higher Spin Theories in AdS(3) and a Gravitational Exclusion Principle,''
  JHEP {\bf 1101}, 142 (2011)
  [arXiv:1012.0598 [hep-th]].
  
\bibitem{other}
M.~R.~Gaberdiel, T.~Hartman,
  ``Symmetries of Holographic Minimal Models,''
  JHEP {\bf 1105}, 031 (2011).
  [arXiv:1101.2910 [hep-th]].
  
  C.~Ahn,
  ``The Large N 't Hooft Limit of Coset Minimal Models,''
  [arXiv:1106.0351 [hep-th]].
  
  M.~R.~Gaberdiel, R.~Gopakumar, T.~Hartman, S.~Raju,
  ``Partition Functions of Holographic Minimal Models,''
  [arXiv:1106.1897 [hep-th]].

  M.~R.~Gaberdiel and C.~Vollenweider,
  ``Minimal Model Holography for SO(2N),''
  arXiv:1106.2634 [hep-th].

  C.~M.~Chang and X.~Yin,
  ``Higher Spin Gravity with Matter in AdS3 and Its CFT Dual,''
  arXiv:1106.2580 [hep-th].
  
\bibitem{Bagchi:2011vr}
  A.~Bagchi, S.~Lal, A.~Saha and B.~Sahoo,
  ``Topologically Massive Higher Spin Gravity,''
  arXiv:1107.0915 [hep-th].
  
\bibitem{Chen:2011vp}
  B.~Chen, J.~Long and J.~b.~Wu,
  ``Spin-3 Topological Massive Gravity,''
  arXiv:1106.5141 [hep-th].
  
  
  
    


\bibitem{Zamolodchikov:1985wn}
  A.~B.~Zamolodchikov,
  ``Infinite Additional Symmetries in Two-Dimensional Conformal Quantum Field Theory,''
  Theor.\ Math.\ Phys.\  {\bf 65}, 1205-1213 (1985).
  


\end{thebibliography}
\end{document}